\documentclass[aps,prl,twocolumn,superscriptaddress,showpacs]{revtex4}

\usepackage{graphicx}
\usepackage{dcolumn} 
\usepackage{bm}      

\newcommand{\Tc}{{T$_c$~}}
\newcommand{\Tn}{{T$_N$}}
\newcommand{\LMO}{{Li$_2$Mn$_2$O$_4$~}}
\newcommand{\usr}{{$\mu$SR~}}

\newcommand{\ub}{{$\mu_{B}$}}

\begin{document}

\preprint{APS/123-QED}

\title{2D Kagom\'{e} Ordering in the 3D Frustrated Spinel \LMO}

\author{C.~R.~Wiebe}
\altaffiliation{Current Address: Department of Physics, Brock
University, St. Catharines, Ontario L2S 3A1, Canada}
\email{cwiebe@brocku.ca} \affiliation{Department of Physics,
Columbia University, New York, New York 10027, USA}
\affiliation{Department of Physics and Astronomy, McMaster
University, Hamilton, Ontario L8S 4M1, Canada}

\author{P.~L.~Russo}
\affiliation{Department of Physics, Columbia University, New York,
New York 10027, USA}

\author{A.~T.~Savici}
\affiliation{Department of Physics, Columbia University, New York,
New York 10027, USA}

\author{Y.~J.~Uemura}
\affiliation{Department of Physics, Columbia University, New York,
New York 10027, USA}

\author{G.~J.~MacDougall}
\affiliation{Department of Physics and Astronomy, McMaster
University, Hamilton, Ontario L8S 4M1, Canada}

\author{G.~M.~Luke}
\affiliation{Department of Physics and Astronomy, McMaster
University, Hamilton, Ontario L8S 4M1, Canada}

\author{S.~Kuchta}
\affiliation{Department of Chemistry, McMaster University,
Hamilton, Ontario L8S 4M1, Canada}

\author{J.~E.~Greedan}
\affiliation{Department of Chemistry, McMaster University,
Hamilton, Ontario L8S 4M1, Canada}

\date{\today}

\begin{abstract}

\usr experiments on the geometrically frustrated spinel oxide,
\LMO, show the development of spin correlations over a range of
length scales with decreasing temperature.  Increased relaxation
below $\sim$ 150 K is consistent with the onset of spin
correlations.  Below 50 K, spin order on a length scale, which is
long range for the \usr probe, appears abruptly in temperature,
consistent with prior neutron diffraction results. The
oscillations in the zero field asymmetry are analyzed using a
three frequency model. By locating the muon site this is shown to
be consistent with the unexpected 2D q = $\sqrt{3}$ x $\sqrt{3}$
structure on the Kagom\'{e} planes proposed originally from
neutron data. Longitudinal field data demonstrate that some spin
dynamics persist even at 2 K.  Thus, a very complex magnetic
ground state, featuring the co-existence of long length scale 2D
ordering and significant spin dynamics, is proposed. This is
unusual considering the 3D topology of the Mn$^{3+}$ spins in this
material.

\end{abstract}

\pacs{75.10.-b, 76.75.+i, 75.40.-s}
\maketitle

\section{\label{sec:level1}Introduction}

The phenomenon of geometric frustration is now seeing a surge of
interest due to the growing number of unique ground states which
arise from networks of spins in triangular motifs.\cite{Greedan},
\cite{Ramirez}  In particular, over the last few years,
discoveries such as heavy-fermion behavior,\cite{Kondo} spin-ice
ordering,\cite{Gingras} and even novel superconductivity
\cite{Hanawa} have been observed in materials which have magnetic
sublattices of corner-shared tetrahedra (such as the spinels and
pyrochlores).  This has led to increased theoretical interest in
these systems, which were originally suggested by Anderson to be
excellent candidates for exotic magnetism, such as the dynamic
resonating valence bond state.\cite{Anderson}  Although very few
examples in the literature exist of this elusive class of
materials, new discoveries of the complex magnetism in these
systems continue to intrigue the condensed matter community.

One of the more curious discoveries of late lies within the spinel
materials Li$_{1+x}$Mn$_2$O$_4$.  These were once targets for the
lithium ion battery community, but they were later discovered to
have complicated magnetic properties based upon the frustrated
sublattice that they have in common.  Complete removal of Li
species results in $\lambda$-MnO$_{2}$, which orders below 32 K
into a magnetic unit cell with 256 spins.\cite{Greedan2} The next
member in the series, LiMn$_{2}$O$_{4}$, exhibits partial
Mn$^{3+}$/Mn$^{4+}$ charge ordering at 280 K followed by a very
complex 3D long range magnetic order below 60 K.  Remarkably, the
long range order co-exists with short range, nearest neighbor
length scale, order down to the lowest temperature investigated,
1.5 K.\cite{Wills1},\cite{Greedan4}  But it is the material
Li$_{2}$Mn$_{2}$O$_{4}$\cite{Wills2} which has perhaps the most
remarkable ground state.  This material is synthesized by chemie
douce, or soft chemistry, insertion of Li ions into
LiMn$_{2}$O$_{4}$.  The magnetic sublattice, which is populated by
Mn$^{3+}$ spins, has a slight tetragonal distortion (due to the
Jahn-Teller effect) from the ideal corner-shared tetrahedral
network that is seen in the cubic pyrochlores. Nonetheless, this
distortion is small, and the sublattice remains, topologically,
three dimensional (see figure 1).  D.C. magnetic susceptibility
data showed signs of short range correlations up to T $\sim$ 400
K.  Neutron diffraction on powder samples revealed broad features
indicative of short range magnetic correlations appearing below
150 K.  With decreasing temperature these features took on a
pronounced Warren line shape, a signature of 2D
correlations.\cite{Warren}  This is a remarkable result given the
3D topology of the Mn sublattice. The intensity of the Warren
reflections increased sharply between 50 K and 40 K in an almost
first-order fashion. The maximum 2D spin correlation length,
obtained by fitting to the Warren function, was $\sim$ 90 \AA, a
relatively large but finite value, which remained unchanged from
40 K to 1.6 K.  From the positions of the magnetic reflections,
the authors suggested that the 2D correlations could be assigned
to a q = $\sqrt{3}$ x $\sqrt{3}$ structure confined to the
Kagom\'{e} sheets (see figure 1), but this assignment was based on
the observation of only two magnetic reflections.\cite{Wills2}

\begin{figure}[t]
 \linespread{1}
 \begin{center}
  \includegraphics[scale=0.32,angle=0]{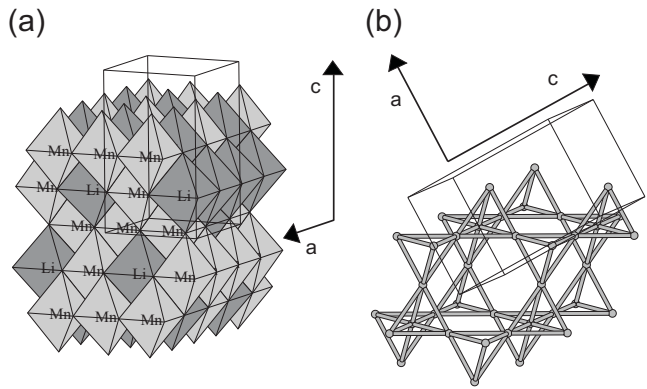}
  \caption{(a)  A polyhedral representation of the structure of
  Li$_{2}$Mn$_{2}$O$_{4}$ showing the Li-O and Mn-O octahedra.
  The Li ions have migrated from the tetrahedral sites in the
  normal spinel, LiMn$_{2}$O$_{4}$, to the octahedral sites in
  Li$_{2}$Mn$_{2}$O$_{4}$.
  (b)  The Mn magnetic sublattice, shown as a network of corner-shared tetrahedra (which would be identical to the cubic pyrochlore sublattice if not for the tetragonal distortion).  This can be
  described as Kagom\'{e} layers which alternate with triangular planar layers in the stacking direction normal to the $<$ 1 1 1$>$ direction.}
  \label{spinel}
 \end{center}
 \linespread{1.6}
\end{figure}

In this paper, $\mu$SR experiments confirm the onset of a long
ranged (on the \usr length scale) spin ordering below 50 K, and
show evidence for the very short range order setting in below 150
K.  Moreover, substantial evidence is provided for the q =
$\sqrt{3}$ x $\sqrt{3}$ magnetic structure on Kagom\'{e} layers
proposed from the neutron data, and may suggest either a 2D order
parameter, or the first-order nature of the transition.

\section{\label{sec:level2}Experimental Procedure and Results}

\subsection{\label{sec:level3}Sample Preparation and Characterization}

\LMO was synthesized according to the method of Wills \textit{et
al.} by chemie douce, or soft chemistry insertion of Li ions into
LiMn$_{2}$O$_{4}$.  LiMn$_{2}$O$_{4}$ was prepared by reacting
stoichiometric amounts of Li$_{2}$CO$_{3}$ and Mn$_{2}$O$_{3}$ in
air at 650$^{o}$ C for 12 hours and 800$^{o}$ C for 24 hours,
followed by a gradual cooling to room temperature. The product was
reground and refired according to the same heating routine and
then tested for phase purity by x-ray diffraction using
K$_{\alpha1}$ radiation on a Bruker D8 diffractometer. The Li
insertion step was completed in an Ar glove box.  Three grams of
finely ground LiMn$_{2}$O$_{4}$ were added to 12.5 mL of 16 M
\textit{n}-butyl lithium in 40 mL of sodium-dried hexane.  This
provides a slight excess of \textit{n}-butyl lithium.  After
gentle heating at 40$^{o}$ C for 5 days, the final product was
filtered and washed well with sodium-dried hexane. The product was
tested for phase purity using a Guinier-H\"{a}gg camera in a
sealed mylar sample holder.

The magnetic susceptibility was measured in a sealed capsule using
a Quantum Design SQUID magnetometer.  The experiments were
completed in field-cooled (FC) and zero field-cooled (ZFC)
sequences and using the reciprocating sample option (RSO) mode on
the instrument. $\mu$SR experiments were completed by enclosing
the sample in a mylar-sealed sample holder with He exchange gas.
All experiments were completed at the M20 beamline at TRIUMF,
Vancouver, Canada in a He-flow cryostat which reached temperatures
from 2 K to 150 K.  Zero-field (ZF) and longitudinal field (LF)
measurements were made in applied fields of up to 2 kG.

\subsection{\label{sec:level4}Experimental Results}

DC magnetic susceptibility measurements made in an applied field
of 500 G qualitatively reproduced the previous results of Wills
and co-workers (see figure 2).  There was no Curie-Weiss behavior
detected up to 350 K, and indeed, previous measurements detected
no such behavior up to 800 K.\cite{Wills2}  This is a signature of
frustrated systems - the Curie-Weiss region is pushed to higher
temperatures due to strong antiferromagnetic interactions between
the spins.  There is also a broad feature at T $\sim$ 110 K which
makes it difficult to perform a Curie-Weiss analysis.  This peak
is interpreted as a signature of short-ranged magnetic ordering of
the Mn$^{3+}$ spins.  A cusp in the susceptibility occurs just
below 50 K, indicative of the onset of the relatively long range
2D order found in the neutron data.  The divergence of the FC/ZFC
data at high temperatures could be due to a small impurity, or the
presence of magnetic correlations.  The lack of impurity lines in
the x-ray diffraction data suggests the latter explanation over
the former.

\begin{figure}[t]
 \linespread{1}
 \begin{center}
  \includegraphics[scale=0.5,angle=0]{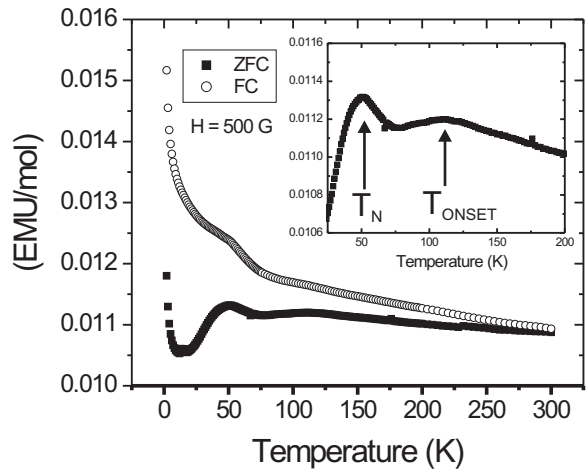}
  \caption{The DC magnetic susceptibility of \LMO for an applied field of H = 500 G.  Zero field-cooled (ZFC) and field-cooled data sets (FC) are denoted.
    A broad feature at $\sim$ 110 K suggests the onset of magnetic correlations (T$_{ONSET}$), and a sharper peak at $\sim$ 50 K corresponds to the N\'{e}el temperature (T$_{N}$).
    The FC/ZFC divergence at higher temperatures is due to magnetic correlations (there is no evidence for impurities through x-ray diffraction measurements).}
  \label{dcchi}
 \end{center}
 \linespread{1.6}
\end{figure}

$\mu$SR measurements in zero-field (ZF) identified a two-component
lineshape which became prominent below 150 K.  This data was fit
to a Kubo-Toyabe function, to account for relaxation processes
from nuclear dipoles, and an exponential function, to account for
relaxation of magnetic origin.

\begin{equation}
P(t) = A(f G_{KT} e^{-\lambda_{slow} t} + (1-f) e^{-\lambda_{fast}
t})
\end{equation}

where P(t) is the muon polarization function, A is the asymmetry,
f is the nuclear dipolar fraction, G$_{KT}$ is the Kubo-Toyabe
lineshape, and $\lambda$$_{slow}$ and $\lambda$$_{fast}$ are the
relaxation rates (see figure 3).  The former rate is slower than
the latter, which accounts for the quasi-static ordering seen
above 50 K as a decrease in polarization at early times.  The
fraction, f, represents the fraction of the slow Kubo-Toyabe
signal of the total signal.  At high temperatures (ie. 150 K),
this is 1.  As one cools down (see figure 5(a)), this fraction
decreases as the fast exponential component grows in size. This
shows that a separate population of spins starts to slow down and
grow in size below 150 K, which correlates nicely with the broad
feature in the DC susceptibility at 110 K.  At 57 K, this
population represents about 25 percent of the total asymmetry.
Below 50 K, about a third of the asymmetry is oscillating
(corresponding to about 50 percent of the spins ordered).

The Kubo-Toyabe lineshape is, explicitly:

\begin{equation}
G_{KT} = \frac{1}{3} + \frac{2}{3} (1 - \Delta^{2} t^{2})
e^{-\Delta^{2} t^{2}/2}
\end{equation}

where $\Delta^{2}$ is the second moment of the field distribution
of nuclear dipolar origin.  A rise in the relaxation rate to a
maximum at 50 K is typical of the dynamics associated with
magnetic transitions to long range ordered ground states.  The
physical origin of this is the longer correlation times near a
phase transition (known as critical slowing down).  The lack of a
sharp maximum at 50 K could be due to either the coarseness of the
temperature steps near \Tn, or the abrupt nature of the
transition.  The appearance of a second component at temperatures
much larger than 50 K, however, is unusual and its origin is due
to the gradual slowing down of spins over decades in temperature
(another common feature of magnetically frustrated systems).  A
likely cause of this is the formation of short-ranged spin
correlations, as suggested by previous neutron work.  It is also
unusual that the relaxation persists to low temperatures below 50
K.  This is evidence for fluctuating spins co-existing with
ordered spins as T $\rightarrow$ 0 K. Further signatures of the
dynamic nature of this ground state is seen in the longitudinal
field measurements (LF), which show that even strong fields do not
completely decouple the signal.

ZF-$\mu$SR spectra taken below 50 K show an oscillating component
from muon precession in a local internal magnetic field.  Fourier
transforms revealed three well-defined frequencies, which were
used to fit the data below 50 K in the regime 0 s $<$ t $<$ 2
$\mu$s (see figure 4).  For longer time scales, the data was fit
to a Kubo-Toyabe lineshape multiplied by an exponential.

\begin{eqnarray}
P(t) = A G_{KT} e^{-\lambda_{slow} t} + A_1 \cos(\omega_1 t)
e^{-\lambda_{1} t} \nonumber\\ + A_2 \cos(\omega_2 t)
e^{-\lambda_{2} t} + A_3 \cos(\omega_3 t) e^{-\lambda_{3} t}
\end{eqnarray}

In this equation, A$_{i}$, $\lambda_{i}$, and $\nu_{i}$ are the
asymmetries, relaxation rates, and frequencies of the three
components, respectively.  A common phase, $\phi$, was used for
the three components for the initial fitting process.  Since this
was within error of zero, the phase was set to zero for the final
fits.

\begin{figure}[t]
 \linespread{1}
 \begin{center}
  \includegraphics[scale=0.35,angle=0]{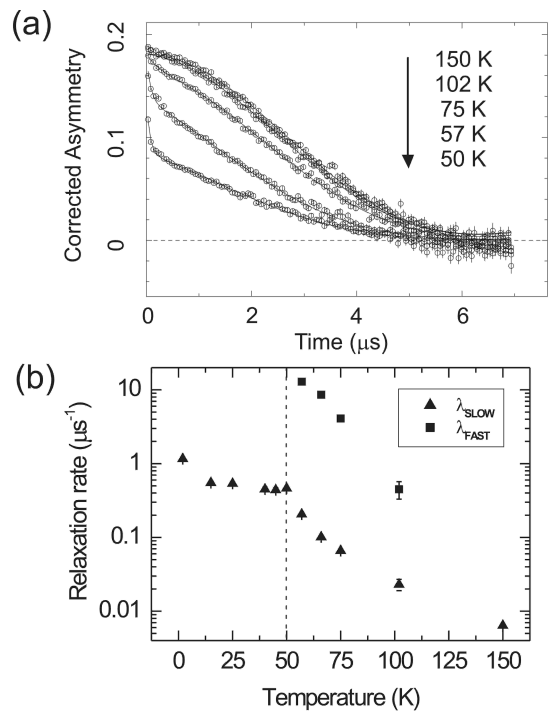}
  \caption{(a) ZF-$\mu$SR spectra at 50 K and above
  with fits described
  in the text.
  (b)  Relaxation rate of the the two components as a function of temperature.  Below 50 K, a different fitting function is
  used for the data.}
  \label{zflambda}
 \end{center}
 \linespread{1.6}
\end{figure}

\begin{figure}[t]
 \linespread{1}
 \begin{center}
  \includegraphics[scale=0.35,angle=0]{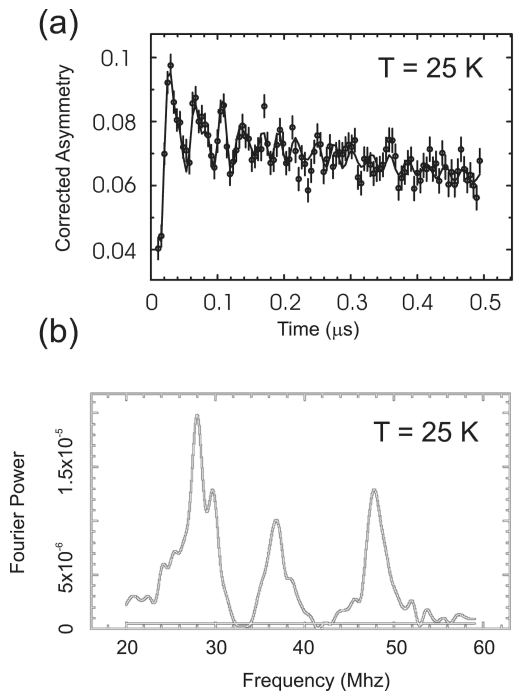}
  \caption{(a)  Early time ZF-$\mu$SR spectra at T = 25 K.  The fit is to a
  three frequency model.
  (b)  The Fourier transform of this data, showing three distinguishable components.}
  \label{fourier}
 \end{center}
 \linespread{1.6}
\end{figure}

\begin{figure}[t]
 \linespread{1}
 \begin{center}
  \includegraphics[scale=0.4,angle=0]{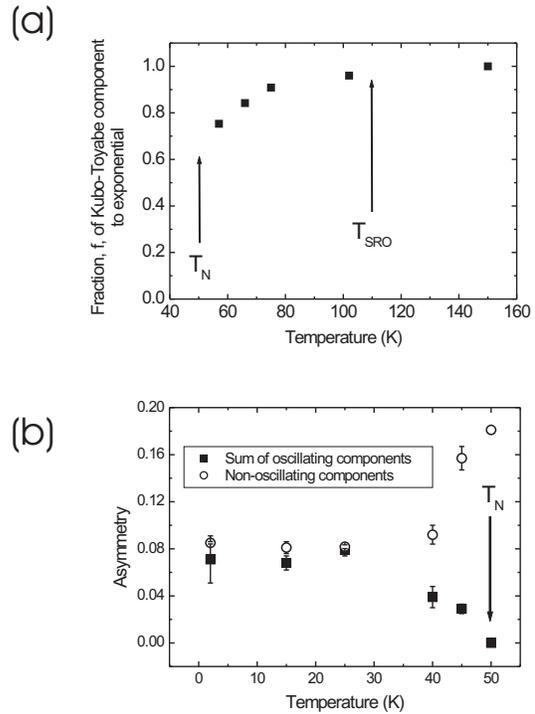}
  \caption{(a)  The temperature dependence of the Kubo-Toyabe
  signal to the exponential signal above 50 K.
  (b)  The temperature dependence of the three oscillating components (summed together) and the
  non-oscillating Kubo-Toyabe function (multiplied by an exponential).}
 \end{center}
 \linespread{1.6}
\end{figure}

\begin{figure}[t]
 \linespread{1}
 \begin{center}
  \includegraphics[scale=0.5,angle=0]{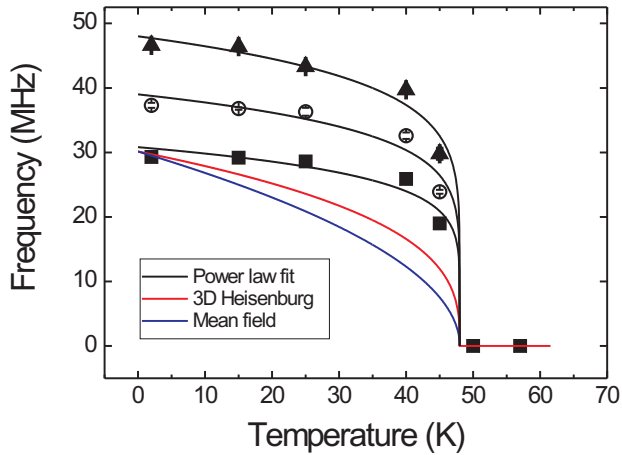}
  \caption{(Colour online) The temperature dependence of the three fitted frequencies.  A power-law exponential has been fitted to the lowest frequency with $\beta$ = 0.14(3).  The two higher frequencies show the same fit for the lower frequency with a normalization factor.  For comparison, the expected power-laws for 3D Heisenburg and mean field theory have been shown.}
  \label{freq}
 \end{center}
 \linespread{1.6}
\end{figure}

\begin{figure}[t]
 \linespread{1}
 \begin{center}
  \includegraphics[scale=0.4,angle=0]{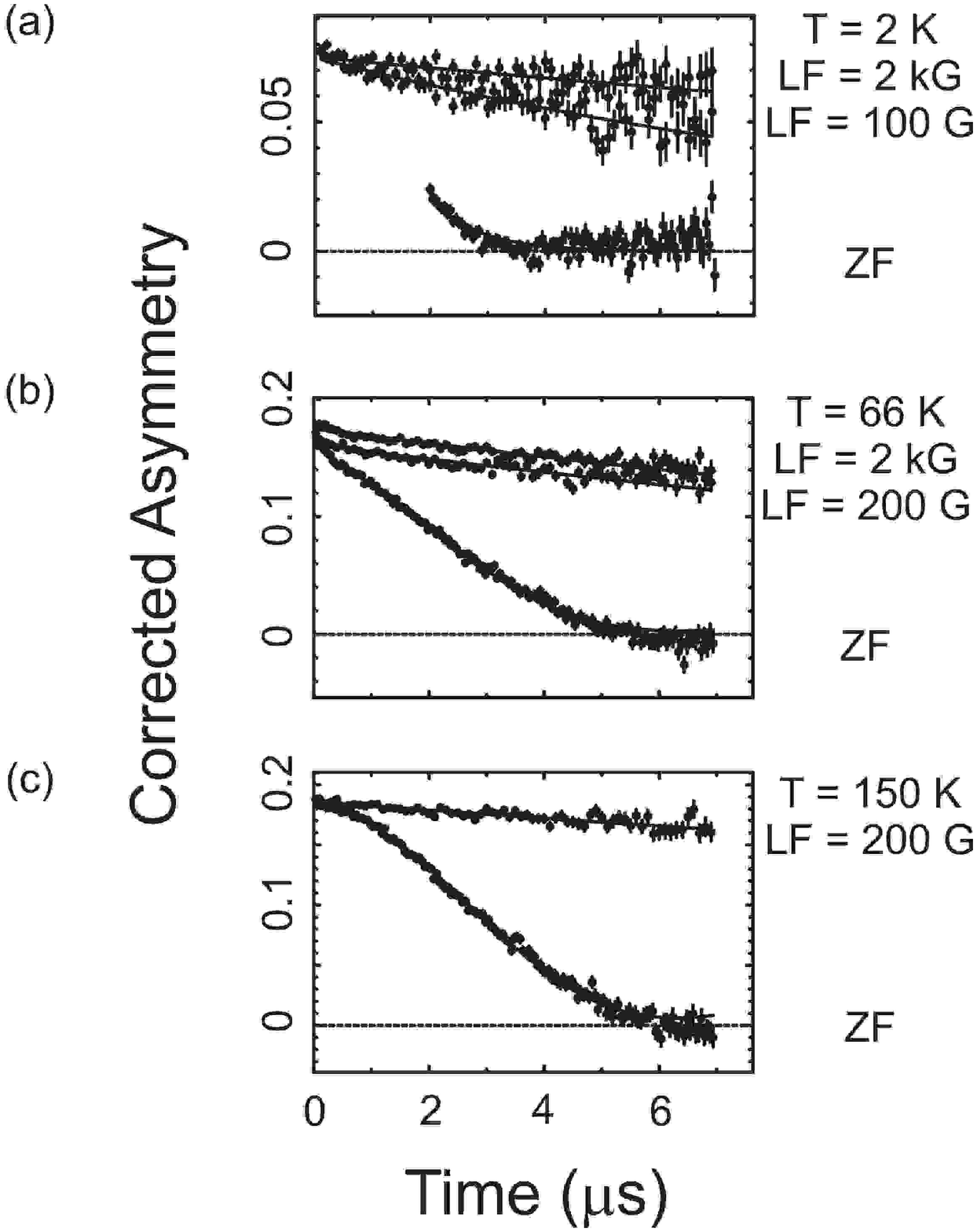}
  \caption{ZF and LF $\mu$SR data at (a) T = 2 K, (b) T = 66 K, (c) T = 150 K.  The data at 2 K in ZF has been omitted in the early time signal where oscillations are observed.}
  \label{lfdata}
 \end{center}
 \linespread{1.6}
\end{figure}

The asymmetries of the oscillating and non-oscillating components
are noted in figure 5.  The drop in asymmetry of the ordered
signal at 50 K coincides with a rise of the
Kubo-Toyabe/exponential signal.

The asymmetry of the oscillating components was found to be only
about a third of the expected total asymmetry, suggesting that
only a portion of the spins are ordered at 2 K.  The remainder of
the spins are either still fluctuating or in short-ranged ordered
spin clusters. The three frequency fit represents the data well,
however the corresponding temperature dependence of all three
components is uncharacteristic of 3D ordering.  All three show a
very rapid increase to saturation below 50 K.

The low frequency component of the data has been fit to the
phenomenological form:

\begin{eqnarray}
\nu = \nu_{0} {(1 - T/T_{N})}^{\beta}
\end{eqnarray}

where $\beta$ is the critical exponent.  Although a rigorous
examination of the power law would require more points around
T$_{N}$, it is clear that the transition is not mean-field like
($\beta$ = 0.5), or 3D Heisenburg ($\beta$ = 0.33), as shown in
figure 6.  The fit, to free parameters for $\beta$, $\nu_{0}$, and
\Tn~ for the lowest frequency component, is $\beta$ = 0.14(3),
which is more consistent with the 2D Ising critical exponent of
$\beta$ = 0.125, and a transition temperature of 48 +/- 2 K. The
rapid increase of the precession frequency is also characteristic
of a first-order like transition, which is echoed in the spin-spin
correlation lengths extracted from neutron scattering
measurements.

The relaxation rate of the 1/T$_1$ tail is approximately a
constant below \Tn.  This is strong evidence that there are
coexisting regions of ordered and disordered spins.  Since the
behaviour of the disordered spins (from the 1/T$_1$ tail) changes
abruptly below \Tn, these spins must be involved with the
transition (it is unlikely that there is bulk phase separation).

LF-$\mu$SR spectra show that most of the signal is decoupled with
the application of applied magnetic fields.  This indicates that
there is a large population of static spins.  However, there is a
small fraction of spins (about 10 percent of the total volume
fraction) which do not decouple, even down to 2 K and in strong
fields (see figure 7 at T = 2 K and T = 66 K). This suggests the
coexistence of magnetic order and regions of persistently
fluctuating moments.

\section{\label{sec:level5}Discussion}

Our $\mu$SR results present strong evidence for the model proposed
by \cite{Wills2}, i.e. a  q = $\sqrt{3}$ x $\sqrt{3}$ ordering for
the Mn spins in the Kagom\'{e} layers.  Below 50 K, there appears
to be a population of spins (roughly 50 percent) which lie in an
ordered state with respect to the muon timescale. In order to shed
some light on the nature of this ordering, one must first attempt
to identify the muon site. Given that there are three separate
frequencies in the data, the first place to start is to assume
that the muon is experiencing three distinct internal fields from
either the magnetic structure or different muon sites or both.  It
is therefore important to locate the muon sites within \LMO.  The
most probable site is one which is equidistant from the oxygen
positions.  There are three obvious choices for this:  at the
interstitial sites (3/8, 3/8, 3/8), (1/4, 1/4, 1/4), or (0, 0.75,
0.125).  Detailed calculations of the field experienced by the
muon at each site were made by using the equation:

\begin{equation}
\Delta^{2}= \frac{4}{9}\gamma_{\mu}^{2} \sum
\overline{\mu_{i}^{2}} \frac{1}{r_{i}^{6}}
\end{equation}

in which $\gamma_{\mu}$ is the muon gyromagnetic ratio (135.5
MHz/T) and $\mu_{i}$ is the dipolar moment at site $i$ at a
distance $r_{i}$ from the muon site.  This is the relaxation rate
$\Delta$ (due to random nuclear dipolar fields) which is extracted
from the Kubo-Toyabe fits to the high temperature data (150 K).
Sums were made over several unit cells in length in the three
crystallographic directions.  The theoretical values for the
relaxation rates at the three sites of interest are listed in
table I.

\begin{table}[htbf]
\begin{center}
\begin{tabular}{|c|c|c|}
\hline
Muon site & $\Delta$ ($\mu$ s$^{-1}$) & Comment\\
\hline
(1/4, 1/4, 3/4) & 0.132 & Occupied by Li \\
(3/8, 3/8, 3/8) & 0.418  & Interstitial  \\
(0, 0.75, 0.125) & 0.308 & Interstitial \\
\hline
\end{tabular}
\caption{Possible muon sites with the calculated relaxation rate.
 The (0, 0.75, 0.125) position has a value which is closest to the fits derived from
 the muon data ($\Delta$ = 0.275 $\mu$s$^{-1}$).}
\end{center}
\label{muonsitetable}
\end{table}

In the material LiMn$_{2}$O$_{4}$, the most probable site is in
the interstitial position (1/4, 1/4, 3/4), where the muon has some
mobility within the lithium-depleted structure acting like a
Li$^{+}$ ion.  This has been verified by detailed $\mu$SR
experiments and calculations on LiMn$_{2}$O$_{4}$.\cite{Ariza}
However, in \LMO, all of these sites are filled by intercalated
lithium atoms. The calculated relaxation rate of 0.132
$\mu$s$^{-1}$ does not agree with the value found from the
Kubo-Toyabe fits of 0.275 $\mu$s$^{-1}$, suggesting that another
muon site is a better choice.

The site (3/8, 3/8, 3/8) is in an electrostatically favorable
environment, surrounded by 6 oxygens in an octahedral framework at
distances of 1.5 to 2.6 \AA.  This position was found to be the
muon site for the spinel LiV$_{2}$O$_{4}$ in a recent $\mu$SR
study.\cite{Koda}  However, the theoretical value for the
relaxation rate from this field distribution is 0.418
$\mu$s$^{-1}$ which does not agree with the value obtained from
our data.

The best agreement with our high temperature Kubo-Toyabe fits to
the relaxation rate of 0.275(1) $\mu$s$^{-1}$ is at the (0, 0.75,
0.125) position.  It is perhaps not surprising that this is the
most probable muon stopping site for \LMO.  This is a tetrahedral
site where Li resides in LiMn$_{2}$O$_{4}$, but it is not occupied
in \LMO according to previous diffraction studies (all of the
lithium atoms are at the octahedrally coordinated \textit{8c} site
(0, 0, 0)).  This is, therefore, a favorable position for the
muon, which will act like a Li$^{+}$ cation surrounded by a
four-fold oxygen cage at distances of 1.92 \AA. It is also
surrounded by 4 equidistant Li$^{+}$ cations at distances of 1.82
\AA.

\begin{figure}[t]
 \linespread{1}
 \begin{center}
  \includegraphics[scale=0.32,angle=0]{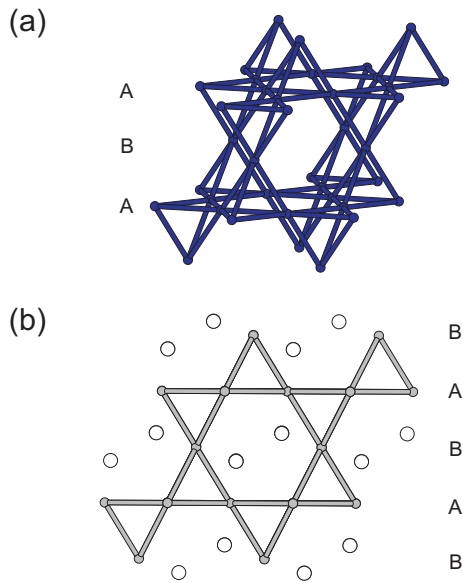}
  \caption{(a)  The Kagom\'{e} magnetic sublattice within the spinel structure.  A and B indicate the alternating Kagom\'{e}
  and triangular planar layer sites, respectively.
  (b)  The muon stopping sites with respect to the Kagom\'{e} (A) and interconnecting layers (B).}
  \label{kagome}
 \end{center}
 \linespread{1.6}
\end{figure}

\begin{figure}[t]
 \linespread{1}
 \begin{center}
  \includegraphics[scale=0.32,angle=0]{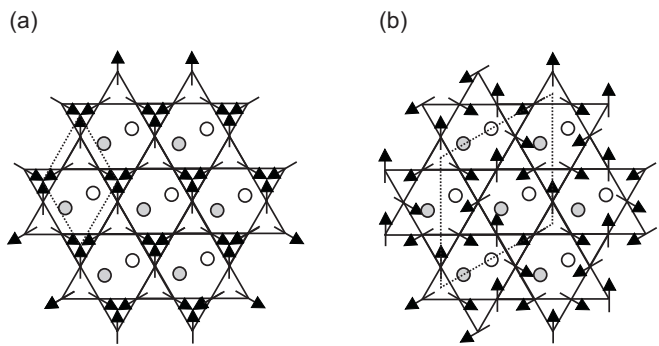}
  \caption{(a)  The q = 0 Kagom\'{e} structure and (b) the q =
  $\sqrt{3}$ x $\sqrt{3}$ structure.  The muon sites in \LMO are
  indicated on the figures as circles, slightly above (grey) and
  below (white) the Kagom\'{e} planes.  The magnetic sublattices
  in 2D are marked by dashed lines.  There are two muon
  sites for the q = 0 structure, but six sites
  for the q = $\sqrt{3}$ x $\sqrt{3}$ structure.}
  \label{kagomestructures}
 \end{center}
 \linespread{1.6}
\end{figure}

Given this result, one can now shed some light on the muon
precession frequencies observed at the ordering transition.  Wills
and co-workers suggested that a q = $\sqrt{3}$ x $\sqrt{3}$
Kagom\'{e} ordering is present within the pyrochlore Mn magnetic
sublattice.  This leaves two separate magnetic species - the
ordered spins which lie within the Kagom\'{e} layers and the
disordered spins in the adjacent triangular plane layers.  Figure
8 shows this explicitly, with the muon stopping sites indicated.
These are located slightly above and below the Kagom\'{e} planes,
at positions roughly in the middle of each Kagom\'{e} plaquette.
Three different muon precession frequencies are seen for our data,
at roughly 46 MHz, 37 MHz, and 30 MHz, corresponding to three
different internal fields seen by the muon.  To see how this could
arise in the q = $\sqrt{3}$ x $\sqrt{3}$ spin structure, consider
figure 9.  The two possible spin structures are plotted with
respect to the muon stopping sites, which lie at positions just
above and below the planes as indicated.  For the q = 0 structure,
the muons all experience virtually the same spin environment
within the magnetic unit cell.  In the upper limit, the muon can
feel two frequencies within the unit cell (one for each stopping
site). However, there are six different internal fields
experienced by muons within the q = $\sqrt{3}$ x $\sqrt{3}$
magnetic unit cell.  Detailed calculations of the internal fields
using dipolar interactions between the moments and the muon sites
reveal that there are indeed three frequencies experienced by the
muons - 4 sites which are nearly degenerate with a low frequency,
and the remaining two sites with higher frequencies.  Table II
shows these results explicitly, with excellent agreement shown
with the ratios of the three frequencies observed in our data.
This also is in qualitative agreement with the distribution of
components in the fourier transform, which reveals a broad set of
frequencies centered about 30 MHz (from the four nearly degenerate
fields), and two smaller components at 37 and 46 MHz (from the
other two fields observed). In conclusion, one can say with some
confidence that the q = $\sqrt{3}$ x $\sqrt{3}$ structure is
consistent with the distribution of frequencies observed in our
data, given our choice of muon site. The q = 0 structure cannot
give rise to three unique muon frequencies.

\begin{table}[htbf]
\begin{center}
\begin{tabular}{|c|c|c|}
\hline
Frequency (MHz) & Ratio (exp.) & Ratio (calc.)\\
\hline
29.3 +/- 0.4 & ---  & ---  \\
37.3 +/- 0.4 & 1.27 +/- 0.04 & 1.34 \\
46.4 +/- 0.4 & 1.58 +/- 0.04 & 1.46 \\
\hline
\end{tabular}
\caption{Experimental frequency ratios for precession components
at 2 K.  The ratios of these frequencies are compared to the
calculated values for the muon site in the $\sqrt{3}$ x $\sqrt{3}$
structure.}
\end{center}
\label{frequencytable}
\end{table}

With this excellent agreement of the internal field at the muon
sites with respect to the ratios of the frequencies observed, one
can now estimate the size of the average static moment at the
Mn$^{3+}$ site.  This is done by evaluating the the size of the
field at these points by dividing the precession frequency by the
muon gyromagnetic ratio ($\gamma_{\mu}$ = 135.5 MHz/T), and then
comparing this to our calculations of the frequency distribution.
The resultant average static moment of 1.5(1) \ub~ per atom is shy
of the expected value for Mn$^{3+}$ moments of 4.0 \ub.  Reduced
moments are common in magnetically frustrated
systems.\cite{Ramirez} Another possible origin of the small moment
is that there are some disordered spins within the ordered
Kagom\'{e} layers, leading to a lower average static moment.

In subsequent LF-\usr experiments, we observed that applied fields
do not completely decouple the signal (see figure 6), which
suggests that the dynamics are not quenched easily.  The rise of
the relaxation rate to a constant value below 50 K is consistent
with a dynamic ground state as well.  Other frustrated systems,
such as Tb$_2$Ti$_2$O$_7$, show this behaviour, which has been
interpreted as a ground state stabilized by quantum
fluctuations.\cite{Gardner}  The asymptotic relaxation rate,
$\sim$ 1 $\mu$s$^{-1}$, is of the same order of magnitude.  This
distinguishes the behavior of this system from that associated
with glassy spin freezing, such as seen in Y$_{2}$Mo$_{2}$O$_{7}$,
which has a prominent cusp in the relaxation rate and a low
temperature value which is about 250 times smaller than
\LMO.\cite{sarah}  Highly correlated dynamics typical of
frustrated materials, such as spin-liquid behavior,\cite{Martin}
would give rise to purely exponential relaxation, or
``undecoupable" gaussian features in zero-field \usr.\cite{Uemura}
\LMO is qualitatively different from either a pure spin glass or a
pure spin liquid, with the co-existence of ordered and fluctuating
spins.

This partial ordering phenomena is rare in magnetism, but perhaps
not unexpected.  Intercalated materials are often found to be
multi-phasic, for example, by virtue of the reaction process,
which depends on the reaction kinetics of butyl-lithium on
micrograins of sample.  This explanation seems unlikely, however,
due to the lack of impurity lines found in x-ray powder
diffraction data using the Guinier camera.  Partial magnetic
ordering at \Tn~ has been seen in a wide variety of materials, and
similar behavior has been seen in the \usr literature.  One
material in particular, the oxygen doped superconductor
La$_2$CuO$_{4+\delta}$, has recently been examined by Savici
\textit{et al.} \cite{Savici}  It is believed, from the
perspective of various experimental probes, that the magnetism
develops with the superconductivity at the same temperature.
However, the magnetic component develops in an abrupt way - that
is, there is a rapid rise in the frequency of the muon signal at
\Tn~ which is atypical of three dimensional phase transitions.
This is not due to a poor quality of fit to the data - like in
\LMO, the frequency has a discontinuous evolution at \Tc.  In
La$_2$CuO$_{4+\delta}$, the explanation for this result is that
there is a spin-density wave structure (as revealed by neutron
scattering) in isolated, two dimensional ``islands'' of spins
which reside within the superconducting fluid.  The small
asymmetry of the oscillating component (of about 30 percent with
respect to the total asymmetry) was used to calculate the the size
of these islands, and a corresponding ``swiss-cheese'' model of
superconductivity was developed to explain the results.  It may be
that in \LMO, similar two dimensional islands of the ordered phase
coexist within a sea of quasi-static spins.  This could be due to
inhomogeneities, or due to some exotic ordering mechanism which
favours domain formation of such a ground state.  Further studies
are clearly needed to distinguish the two possibilities.

With respect to other spinel oxides, \LMO has common features with
materials such as ZnFe$_2$O$_4$.  The low temperature ground state
of ZnFe$_{2}$O$_{4}$ is an ordered antiferromagnet with a k = (0,
0, 0) wavevector at 10 K.\cite{znfe2o4}  However, it has been
known for some time that a short range ordered state exists at
temperatures much higher than \Tn~ - as high as 100 K.  Broad
peaks appear in elastic neutron scattering experiments which
persist to low temperatures. In addition, the \usr data on this
material shows a two-component lineshape below 100 K in accordance
with the slowing down of spins in a short range ordered state.
With neutron and \usr studies combined, the consensus is that at
\Tn = 10 K, only about 30 percent of the volume fraction is long
range ordered. The remaining fraction are in
``superantiferromagnetic'' clusters of short range ordered spins
with sizes of about 30 \AA~ (incidentally, this is in good
agreement with the spin-spin correlation length in the \LMO
neutron scattering data of about 20 \AA~ at 100 K \cite{Wills2}).
The similarities between these two materials are considerable,
both with short range ordered states at high temperatures,
followed by a transition to a co-existing long range ordered state
at low temperatures.  The \usr data on ZnFe$_2$O$_4$ suggests
there is some competition between these two states below \Tn, with
the volume fraction of the LRO clusters becoming larger at the
expense of the SRO clusters.  This has not been observed in \LMO.
Larger fields are also needed to decouple the LF-\usr data in
ZnFe$_2$O$_4$, which suggests that significant dynamics play a
role at low temperatures. The ground state is inhomogeneous in
both cases, but notable differences exist between these two
materials.

A closer relative to \LMO is the single-lithiated spinel
LiMn$_2$O$_4$. As mentioned above, this material orders into a LRO
state at 60 K. The magnetic unit cell is enormously complicated,
with a k = (1/2, 1/2, 1/4) structure and 1152 spins
within.\cite{Wills1} A detailed magnetic structure is at this time
unavailable. However, it is known that there is a SRO state which
exists to high temperatures.  Recently, spin polarized neutron
scattering measurements revealed that above \Tn = 65 K the
observable magnetic cross section is entirely due to SRO spins on
a length scale of $\sim$ 3.5 \AA (second Mn neighbors).
\cite{Greedan4} As the temperature is reduced below \Tn, the
fraction of the total magnetic cross section associated with LRO
increases, reaching a maximum value of only 50 percent at 1.5 K.
Inelastic scattering shows that even at 1.5 K, $\sim$ 20 percent
of the spins are still fluctuating.  LiMn$_{2}$O$_{4}$ thus shows
a complex magnetic ground state with co-existing 3D LRO, SRO, and
fluctuating spins. This is similar to the situation found for \LMO
but the details are different as there is no true LRO in this
material but only 2D SRO of significant, but finite, extent at the
lowest temperatures.

In broader terms of comparison, \LMO is very distinct from other
frustrated sublattices.  Gd$_2$Ti$_2$O$_7$ has been suggested to
be an example of a system which has a Kagom\'{e} ordering on a
pyrochlore lattice.\cite{champion}  The mechanism is very
different in this case, with dipolar interactions influencing the
low-temperature physics of the Gd moments ordering.  However, the
real canonical examples of the 2D Kagom\'{e} lattice are the
jarosites.  For years, SrCr$_{x}$Ga$_{12-x}$O$_{19}$ (SCGO) was
considered to be an example of the unusual physics which develops
in Kagom\'{e} systems, being a material which has exhibited
spin-glass to spin-liquid characteristics depending on the Cr-site
coverage and sample quality.\cite{Uemura},\cite{Ramirez2} However,
this material, being of the magnetoplumbite structure, is better
described as an array of bilayer Kagom\'{e} units rather than the
pure structure. The jarosites, of general formula
AB$_3$(SO$_4$)$_2$(OH)$_6$, with B being the magnetic species,
represent the best physical realization of the Kagom\'{e}
lattice.\cite{Townsend}, \cite{Wills3} Although disorder within
these layers is common, most of these materials have a high site
coverage that is still within the percolation limit.  A large
portion of them form LRO ground states as well, in either the q =
0 or q =$\sqrt{3}$ x $\sqrt{3}$ structures.\cite{Keren}  The
latter structure has been shown to be more stable than the former
due to quantum effects which choose this state from the multitude
of classically degenerate states.\cite{Sachdev}  However, if one
includes next nearest-neighbor interactions, the situation becomes
more complicated.  For second neighbor (J$_{2}$) and third
neighbor interactions (J$_{3}$), the q = 0 state is preferred for
J$_{2}$ $>$ J$_{3}$ and the q =$\sqrt{3}$ x $\sqrt{3}$ for J$_{2}$
$<$ J$_{3}$.\cite{Harris}  The situation becomes even more
elaborate if interplane interactions are considered as
well.\cite{Wills4} More detailed studies are needed to clarify the
nature of the spin interactions in \LMO, such as been done
recently for the jarosite KCr$_{3}${OD}$_{6}$(SO$_{4}$)$_{2}$ with
inelastic neutron scattering experiments.\cite{Lee}

It is unclear why a three dimensional system such as \LMO would
order with small regions of correlated 2D spins. Although these
spin clusters have been observed in other spinels, as mentioned
above, there is as of yet no explanation for why this would occur.
The most intriguing work as of late is on the spinel
ZnCr$_{2}$O$_{4}$, in which the authors, through neutron
scattering experiments, demonstrate that hexagonal, 2D spin
clusters play an important role in the low-temperature spin
dynamics.\cite{Lee2} These hexagonal ``protectorates'' form
independently of one another, and their excitations are the origin
for the local zero energy modes in the pyrochlore lattice.  It is
suggested that the formation of spin clusters might be a central
theme of self-organization in frustrated materials, which in many
cases cannot have conventional N\'{e}el ground states.  Future
work to measure the inelastic spectra of \LMO would be essential
in elucidating the physics of these many-body systems.

\begin{acknowledgments}

C.~R.~Wiebe would like to acknowledge support from NSERC in the
form of a PDF.  The authors would like to thank the financial
support of NSERC and CIAR.  The authors are also grateful for the
technical support of Bassam Hitti and Don Arseneau at TRIUMF.

\end{acknowledgments}

\end{document}